\documentclass[sigconf,authorversion,nonacm]{acmart}

\usepackage{lipsum}

\newcommand\blfootnote[1]{%
  \begingroup
  \renewcommand\thefootnote{}\footnote{#1}%
  \addtocounter{footnote}{-1}%
  \endgroup
}

\AtBeginDocument{%
  \providecommand\BibTeX{{%
    \normalfont B\kern-0.5em{\scshape i\kern-0.25em b}\kern-0.8em\TeX}}}

\usepackage[utf8]{inputenc}
\usepackage{verbatim}
\usepackage{amsmath}
\usepackage{graphicx}
\usepackage{subfigure}
\usepackage{multirow}
\usepackage{tabularx}
\usepackage{comment}
\usepackage{xspace}
\usepackage{makecell} 
\usepackage{algorithm}
\usepackage{algpseudocode}
\usepackage{paralist, tabularx}

\usepackage{tikz}

\usepackage{enumitem}

\newcommand{\Nseven}{\textit{Medium}}
\newcommand{\Nsix}{\textit{Large}}
\makeatletter

\usepackage{amsfonts}
\usepackage{bbm}
\usepackage{booktabs}
\usepackage{svrsymbols}
\usepackage{mathtools}

\usepackage[normalem]{ulem}

\algdef{SE}[DOWHILE]{Do}{doWhile}{\algorithmicdo}[1]{\algorithmicwhile\ #1}%
\algnewcommand{\algorithmicforeach}{\textbf{for each}}
\algnewcommand\algorithmicbreak{\textbf{break}}
\algnewcommand\Break{\algorithmicbreak}
\algdef{SE}[FOR]{ForEach}{EndForEach}[1]
  {\algorithmicforeach\ #1\ \algorithmicdo}
  {\algorithmicend\ \algorithmicforeach}
\algtext*{EndIf}
\algtext*{EndForEach}
\algtext*{EndWhile}

\acmConference{1st GNNet Workshop at CoNEXT'22}{December, 2022}{Rome, Italy}

\title[Cross-network transferable interference estimation]{Cross-network transferable neural models\\ for WLAN interference estimation}

\author{Danilo Marinho Fernandes}
\affiliation{
   \institution{Huawei Technologies Co. Ltd} \institution{Institut Polytechnique de Paris}
    \country{France}}  
\email{danilo.marinho-fernandes@polytechnique.edu}

\author{Jonatan Krolikowski, Zied Ben Houidi, Fuxing Chen, Dario Rossi}
\affiliation{%
  \institution{Huawei Technologies Co. Ltd}
  \country{France}}
\email{jonatan.krolikowski, zied.ben.houidi,}
\email{chenfuxing, dario.rossi@huawei.com}

\acmYear{2022}\copyrightyear{2022}
\setcopyright{acmcopyright}
\acmConference[GNNet '22]{Graph Neural Networking Workshop}{December 9, 2022}{Roma, Italy}
\acmBooktitle{Graph Neural Networking Workshop (GNNet '22), December 9, 2022, Roma, Italy}
\acmPrice{15.00}
\acmDOI{10.1145/3565473.3569184}
\acmISBN{978-1-4503-9933-3/22/12}

\begin{abstract}
Airtime interference is a key performance indicator for WLANs, measuring, for a given time period, the percentage of time during which a node is forced to wait for other transmissions before to transmitting or receiving. Being able to accurately estimate interference resulting from a given state change (e.g., channel, bandwidth, power) would allow a better control of WLAN resources, assessing the impact of a given configuration before actually implementing it. 

In this paper, we adopt a principled approach to interference estimation in WLANs. We first use real data to characterize the factors that impact it, and derive a set of relevant synthetic workloads for a controlled comparison of various deep learning architectures in terms of accuracy, generalization and robustness to outlier data. We find, unsurprisingly, that Graph Convolutional Networks (GCNs) yield the best performance overall, leveraging the graph structure inherent to campus WLANs. We notice that, unlike e.g. LSTMs, they struggle to learn the behavior of specific nodes, unless given the node indexes in addition. We finally verify GCN model generalization capabilities, by applying trained models on operational deployments unseen at training time.
\end{abstract}

\begin{document}

\renewcommand{\shortauthors}{Fernandes et al.}
\maketitle

\section{Introduction}
\blfootnote{DOI: 10.1145/3565473.3569184}
Wireless Local Area Networks (WLANs) are today the preferred access technology to connect users of campus and enterprise networks through a number of wireless Access Points (APs). 
WLAN radio resources are shared among users (or: STAs for stations) and each AP is configured to use a specific channel in a half-duplex fashion, with CSMA/CA listen-before-talk mechanism to avoid \emph{physical} interference \emph{(collisions)}.  Whenever a device $a$ intends to transmit, it first detects if the channel is occupied by other transmissions from \emph{neighboring} transmitters. If that is the case, $a$ backs off and waits for a randomly chosen time before trying again. We refer to the waiting time that $a$ incurs as \emph{airtime interference}. 
The main factors that determine airtime interference are thus (i) the channel configuration of transmitters, which dictates which nodes potentially transmit on the same channel, (ii)  the neighborhood relationship between transmitters, i.e. which nodes are detectable when transmitting on the same channel and finally (iii) the transmit duration of interfering nodes. 
High interference is undesirable as it leads to high delays and low throughput. 
Accordingly, radio resource management strategies aim at interference mitigation 
by acting on channel and power configuration: being able to accurately predict the effect of resource re-configuration is thus desirable for optimal WLAN control. Unfortunately though,  accurate \textit{interference estimation} is still challenging~\cite{iacoboaiea2021real,shrivastava2011pie}.

In this paper, we define our interference estimation problem,
inline with the recent trend of using data-driven what-if scenario simulators~\cite{rusek2020routenet,soto2021atari}. There, the idea is to build data-driven models that predict the value of some metrics assuming some hypothetically simulated given input. As far as interference is concerned, accurately knowing the interference levels that result from a given hypothetical network configuration allows to search for the best configuration that minimizes interference, or to know beforehand the impact of an action before implementing it in the real network.
While it is possible to estimate the AP loads resulting from a configuration and some user demands, knowing the ensuing interference levels at each AP is more challenging. 


Tackling this problem, we adopt in this paper a principled approach to WLAN airtime interference estimation.
First, we analyze in Sec.~\ref{sec:overview}  real data collected from two production networks with thousands of users, to systematically characterize the factors that impact interference as a function of AP loads and topology. 
We then consider a set of neural models for its estimation in Sec.~\ref{sec:models}.
Guided by the impacting factors earlier identified, we design in Sec.~\ref{sec:synthetic} a synthetic benchmark allowing us to compare different types of neural networks in terms of (i) accuracy, (ii) generalization to variable size networks as well as the (iii) ability to capture the specific behavior of certain nodes. Finally, we illustrate the predictive and transfer power of trained models by testing them on different operational networks in Sec.~\ref{sec:real}. We conclude by discussing related work in Sec.~\ref{sec:related} and summarizing our findings in Sec.~\ref{sec:conclusions}.

\section{Interference in the wild}
\label{sec:overview}

\subsection{Datasets}
As tabulated in Tab.~\ref{sec:data:networks}, we use two datasets, collected respectively in a 5GHz medium size network of 33 APs (\Nseven) measured during 6-months, and a larger but less loaded 84 AP network (\Nsix) measured during one month. We use the first to characterize interference and train neural models and the second to test the transfer abilities of the models. 
At a glance, we remark that the two selected networks are significantly heterogeneous: in terms of size/density, as well as in terms of load. This choice is done on purpose to stress test limit of transferability of learned models, as we shall see later.

These networks periodically collect telemetry at different timescales. In this paper 
we limit our attention to 10-minutes samples of the following AP-level features:
\begin{itemize}[leftmargin=*]
\item 
\textit{tx time}: \% of airtime used for transmissions \emph{from} the AP during a sample period.
\item \textit{rx time}: \% of airtime of transmissions \emph{towards} the AP.
\item \textit{interference}: \% of airtime interference for an AP.
\item \textit{topology}: Average received signal strength indicator (RSSI) toward/from each other AP.
\end{itemize}
We define the \emph{load} of an AP as the sum of its tx and rx time, i.e.\ the \% of airtime used by the AP itself.

\subsection{Model-based estimation baselines}
We now devise two simple baseline estimators that will help with the data characterization.
\subsubsection{Simple sum} As a first naive baseline, we use a simple predictor~\cite{iacoboaiea2021real} that is based on following assumptions. First, all STAs associated to an AP experience  the same radio conditions as the AP; this allows to consider the STA tx time in sum the same as the associated AP rx  time without further knowledge about specific STAs. 
Second, same-channel APs sense each other's transmissions whenever the RSSI exceeds a sensing threshold: such APs are considered \emph{neighbors}, and  cannot transmit simultaneously. Third, for the sake of simplicity we assume there is no interference from external sources (such as third-party APs, radar signals etc).
Based on the above assumptions, the airtime interference can be estimated as a  \emph{simple sum} of the load times of all same-channel neighbors:
$
I^{\text{ss}}_{a} = \sum_{b\in\text{neigh}(a)} \ell_{b},
$
where $\text{neigh}(a)$ are the same-channel neighbors of AP $a$ and $\ell_{a}\in[0,1]$ is the load.

\subsubsection{Uniform superposition}  Since, in reality, two neighbors may transmit at the same time (\emph{concurrent transmission}), we introduce a probabilistic refinement to account for traffic burstiness in concurrent transmissions of neighbors,  which we refer to as \textit{uniform superposition}. Here, the transmissions times of all neighbors are spread uniformly randomly and independently across the time period. This leads to an expected interference of
\begin{equation}
I^{\text{us}}_{a} = 1-\Pi_{b\in\text{neigh}(a)} (1-\ell_{b}).
\end{equation}
\noindent that can compensate the above overestimation. 

\begin{table}[!t]
\caption{Summary of dataset properties.}\label{tab:datasets}
\scriptsize
\begin{tabular}{cccccc}
\toprule
\bf Network    & \bf Duration (Period)  & \bf Samples & \bf AP & \bf  Usage$^\ast$ & \bf Purpose \\ 
\midrule
\Nseven  & 176d  (Aug'21-Jun'22)   &  100k     & 33 & 40\% & Train/Val  \\ 
\Nsix   &   22d (Feb-Mar'22)    & 10k     & 84 &  20\% & Test  \\    
\bottomrule 
\multicolumn{5}{l}{$\ast$ Samples where interference is $\geq 10\%$}
\end{tabular}
\label{sec:data:networks}
\end{table}



\begin{figure}[t]
\centering
\includegraphics[trim=0pt 0pt 0pt 0pt, clip, width=1\columnwidth]{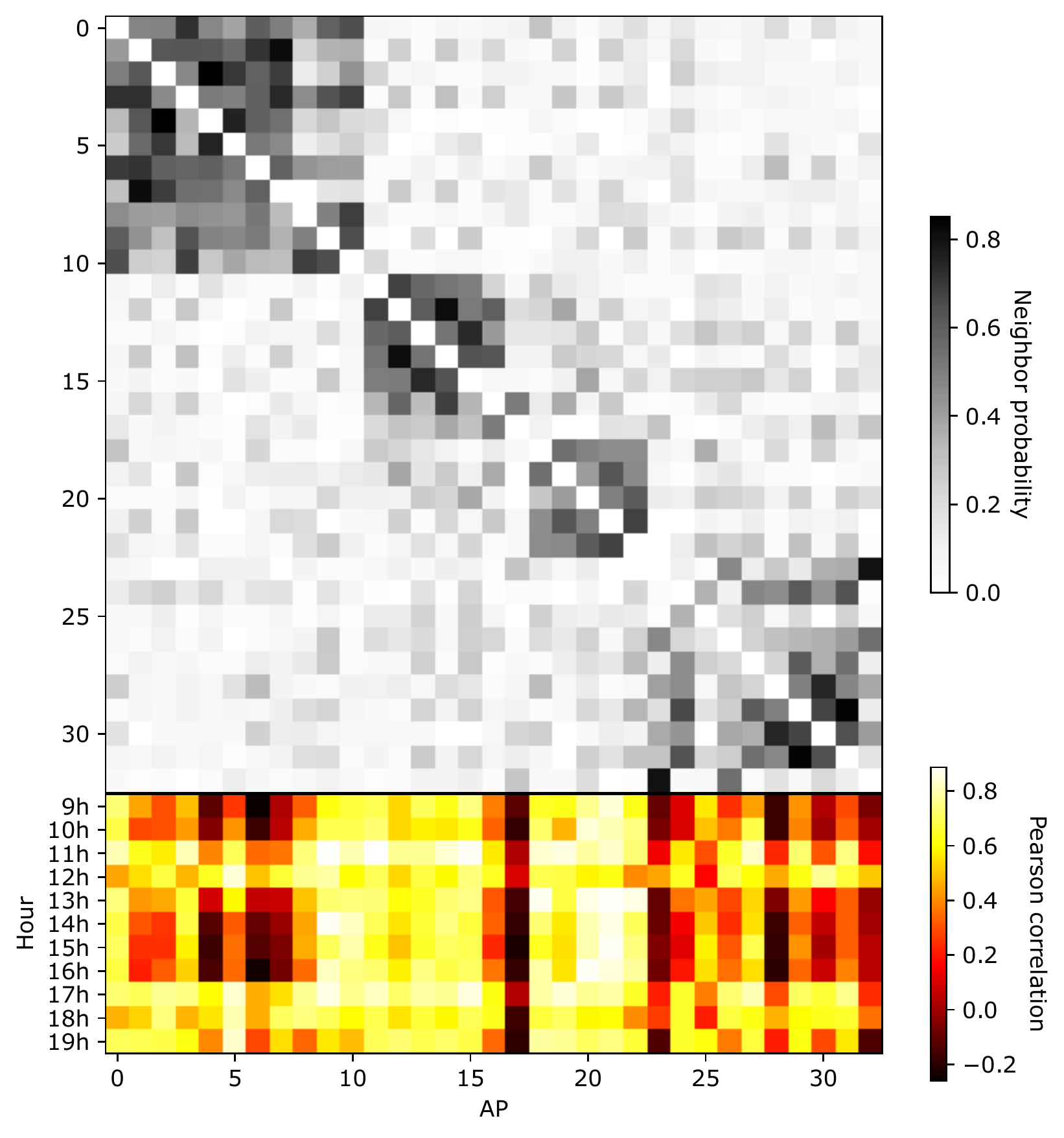}
\caption{Heatmap of neighborhood probability among all AP pairs (top); correlation between simple-sum and groundtruth across hours of the day (bottom).}\label{fig:combined_heatmap}
\end{figure}

\subsection{Data characterization}
We aim at improving over the aforementioned simple estimators with data-driven models. In order to properly design our solution, we first characterize our datasets.
\subsubsection{Topology} We characterize WLAN topology, as typically done, 
by assuming that two APs are neighbors if the RSSI exceeds the clear channel assessment (CCA) threshold, usually set to -82dB.  Since network configurations (power and channel)  and environment change over time, the neighborhood relationship between AP pairs  fluctuates.
We capture this effect, for the \Nseven\ network, in Fig.~\ref{fig:combined_heatmap} (top) which shows the probability of neighbor relations for all pairs. To ease readability, we apply the Louvain community detection algorithm~\cite{blondel2008fast} on the probability matrix and sort the APs accordingly. We observe that (i) few communities  with more frequent neighborhood relationship are clearly visible,  (ii) even within a community, APs have varying neighborhood relationship and that (iii) the heatmap is noticeably noisy, i.e. many neighbor relations neither always nor never present, providing for a variety of topologies over time.

\subsubsection{Topology impact on interference estimation}  
\begin{figure}[t]
\centering
\includegraphics[trim=10pt 10pt 10pt 10pt, clip,width=1\columnwidth]{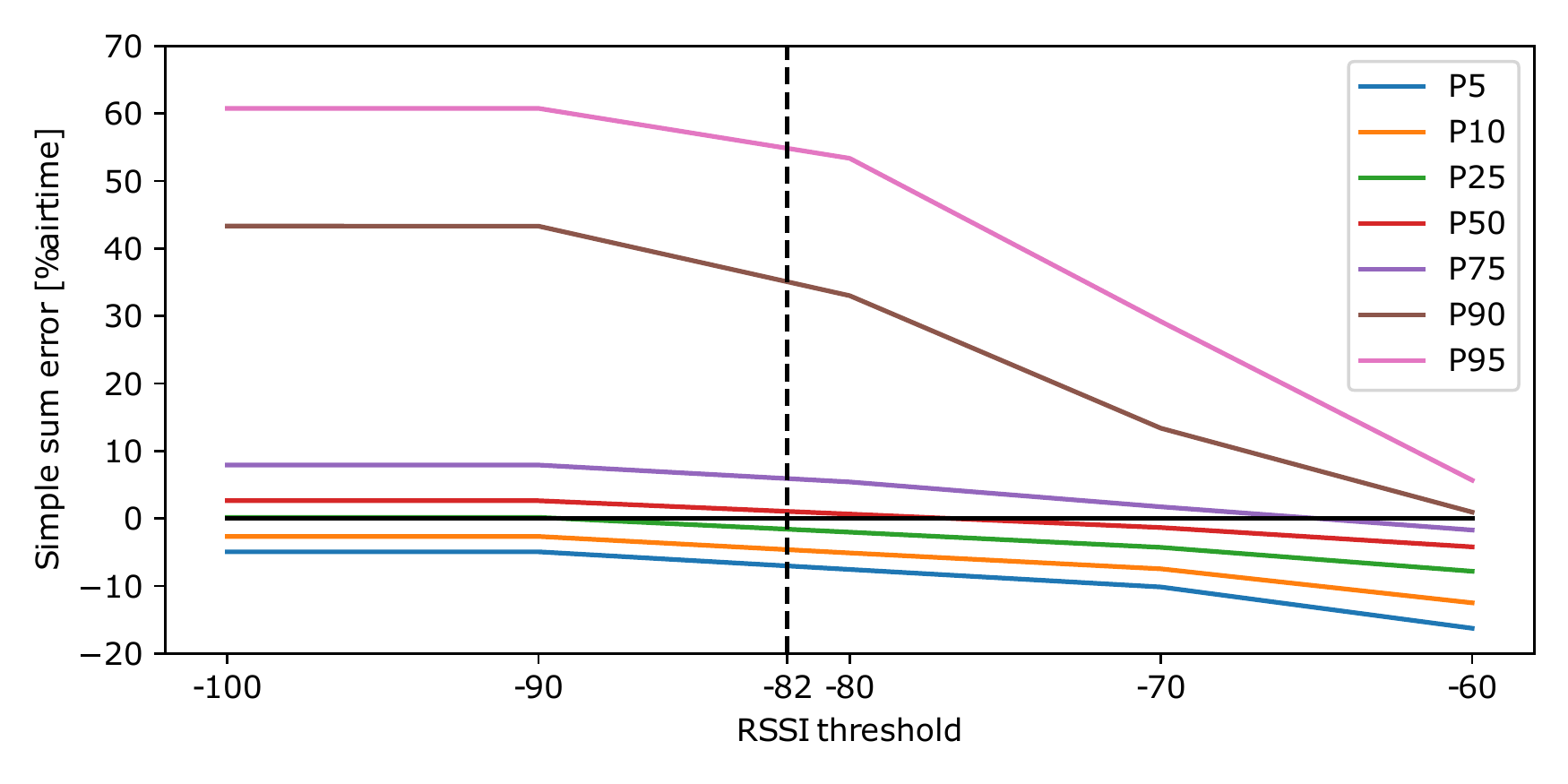}
\caption{Simple sum error over RSSI thresholds}\label{fig:ss_error_threshold}
\end{figure}

We further investigate the effect of topology variability on the interference estimation error, leveraging the simple sum model for the sake of simplicity.
 We let the CCA RSSI threshold vary in  a larger range of values between between -100\,dBm (the minimum reported value) and -62\,dBm  (the typical value for energy detection threshold), which affects the definition of neighbor relationship.  
 Fig.~\ref{fig:ss_error_threshold} reports the percentile of the airtime interference estimation error as a function of the neighborhood threshold.
 Particularly, we see that while the majority of errors are small (median error is negligible and 75\% of the errors are  below 10\%) however a non neglibigle fraction of estimates  (5\% to 10\%) exhibit significant error (in excess of 50\% or 40\% respectively). Correcting such large yet relatively rare errors can hardly be done with a model based approach,  motivating the need for data-driven approaches that we explore in this paper.
 
\subsubsection{Spatio-temporal variability of interference estimation} 
We next revert to the classic CCA threshold, and evaluate variability across space and time of model-based estimation.
Fig.~\ref{fig:combined_heatmap} (bottom) reports a spatio-temporal heatmap of the Pearson correlation coefficient between the measured interference and the interference estimated by simple sum model for different time of day (y-axis) and APs (x-axis).
The heatmap clearly shows that for several APs the simple-sum model provides a consistently good estimation over time (e.g., APs 9--16 and 18--22 exhibit a Pearson correlation $\geq 0.7$  with  $p \leq 0.05$). Conversely, others APs apparently show 
an erratic behavior, with correlation that can be seldom (AP 4, 6, 27, 32) or consistently (AP 17) off with respect to estimation of the simple-sum model (i.e.,  $\leq 0$  with  $p \leq 0.05$). These observations clearly raise the need to account for very specific (and possibly anomalous) behaviors, which
motivates the design of adversarial experiments (Sec.~\ref{sec:singleAP:failure}).

%
\section{Neural models}
\label{sec:models}
We set out to design data driven models that can exploit adjacency relations between neighbors and learn the typical temporal correlation between transmissions, namely 
a multi-layer perceptron (MLP), a bidirectional LSTM, and a Graph Convolutional Network (GCN) \cite{kipf2016semi}. 
Each model is trained with shuffled samples from a specific period, and validated on a different period. All models  employ ReLU activation, and use the Mean Squared Error (MSE) between the predicted interference vector and the groundtruth as loss function. 

\subsection{Baseline neural models}
\subsubsection{MLP} As a baseline neural model, we consider an MLP with 3 combinations of a hidden dense (3000 units) and dropout($0.5$) layer. Inputs are the load vector concatenated with the flattened adjacency matrix.

\subsubsection{LSTM}  As a second standard neural baseline, we consider a recurrent model comprising 3 bidirectional LSTM layers, each with 40 units and each followed by a Dropout($0.5$) layer, and a final Time Distributed Dense layer. The input consists of a sequence of timesteps, one per node\footnote{For example $33$ timesteps during training on the \Nseven\ network and $84$ when testing on the \Nsix\ one.}, where the $i^{th}$ timestep receives the list of all loads concatenated with the $i^{th}$ row of the adjacency matrix.

\subsection{Graph convolutional model}
\label{sec:GCN}
We finally devise GCN models, that we describe in more detail.
When describing model inputs, we adopt the following terminology. The \textit{RSSI matrix} refers to the matrix with the RSSI values between APs. By \textit{adjacency matrix} we denote the thresholded (-82 dB) RSSI matrix with ones for neighbors and zeros otherwise. The interest of using the raw RSSI values comes form the fact that there may exist falsely predicted neighbors, which is more likely for smaller RSSI values. This offers the models the possibility to ``learn the neighborhood threshold''. With these definitions in mind, we first remind the propagation rule for  GCNs: 
\begin{equation} H^{(l+1)} = \sigma(TH^{(l)}W^{(l)}) 
\label{GCN_propagation}\end{equation}


Here, $H^{(l)} \in \mathbb{R}^{n \times d^{(l)}}$ is the 
input of the $l^{th}$ layer, where $n$ is the number of nodes, $d^{(l)}$ is the features dimension per node, 
$\sigma$ is an activation function, $W^{(l)}$ is a layer-specific trainable weights matrix and T is a function of the adjacency matrix called the \emph{kernel}. Instead of setting T as the normalized adjacency matrix with added self-connections and using the propagation rule \eqref{GCN_propagation} as originally proposed~\cite{kipf2016semi}, we employ a simple extension leveraging~\cite{verma}.
 We define 3 kernels $T_1$, $T_2$ and $T_3$, where $T_1$ is the simple identity matrix, and $T_2$, $T_3$ are respectively the RSSI matrix and the  adjacency matrix, both with added self-connections and normalized to prevent numerical instabilities. We then use the following propagation rule, where we have, for a general number of kernels $k$,
\begin{equation} H^{(l+1)} = \sigma(I^{(l)}W^{(l)}) \end{equation}
with
\begin{equation} I^{(l)} = [I_1^{(l)} \cdots I_k^{(l)}] = [T_1H^{(l)} \cdots T_kH^{(l)}] \end{equation}
where $I^{(l)} \in \mathbb{R}^{n \times kd^{(l)}}$ and $W^{(l)} \in \mathbb{R}^{kd^{(l)} \times d^{(l+1)}}$, so that the output dimension per node is $d^{(l+1)}$. This way, we have a weight-shared propagation for all nodes, which can depend on the ensemble of kernels in non-trivial ways.


We leverage real RSSI measurements from the \Nseven\ network to empirically evaluate in Sec.~\ref{sec:kernels}    the advantages of adding   kernel $T_3$.
We use a GCN with 4 graph convolution layers, each having 100 units, that is, $d^{(1)}, d^{(2)}, d^{(3)}, d^{(4)} = 100$, $ d^{(5)} = 1$ and $d^{(0)}=1$ or $1+n$, depending if node IDs are used. Using multiple convolutions allows to use distant neighbor information as well as to learn representations that capture structural equivalence between different nodes. In our data, this leads an average error reduction of $10\%$ compared to a single convolution.

\section{Model Comparison and Tuning}
\label{sec:synthetic}
In order to find the best model for predicting interference on real networks, we design a set of synthetic scenarios on which we analyze the performance of various models.
\subsection{Scenarios}
We design two families of scenarios to validate model learning abilities in scenarios with controlled noise/anomalies. We randomly generate samples of undirected Erdős–Rényi graphs $G(n,p)$ with  a probability of a given pair of nodes being connected of $p=0.2$. For each graph, we assign a feature vector of dimension $(n,d)$, where each row corresponds to the $d$-dimensional feature vector of a given node. We then try to predict scalar node labels, given by particular functions that we define for each case.
In both scenarios, we take respectively $6000$ and $2000$ samples for training and validation. In order to evaluate how well different algorithms generalize for unseen topologies, we randomly generate a set of $k$ fixed topologies, from which graphs of the training set are sampled uniformly at random. The validation set may take any of the $2^{n \choose 2}$ existing topologies. We then analyze how prediction errors change as $k$ increases, by taking the mean squared error (MSE) across the set of all nodes of all validation samples, averaged over $30$ experiments for each $k$ in order to avoid bias due to specific topology choices.

\subsubsection{Simple-sum toy case}\label{sec:simplesum}
As the toycase \emph{simple-sum problem},  we predict node labels defined as the sum of one-hop neighbor features. Here, labels are scalars assigned as integers uniformly sampled in [0.01,1], representing AP self-utiliziation. This problem could be trivially solved by a GCN with a single convolution, but serves as a baseline to evaluate how well other architectures can generalize when trained with a fixed set of topologies.

\subsubsection{Single AP failure}\label{sec:singleAP:failure}
We design a second scenario adding an AP with erratic (but controlled) behavior, as we observed earlier in real data. Essentially, all nodes behave the same way as in the simple-sum case, except for a particular node with fixed index that always has label 0. The goal is to judge the models's capacity to  identify the anomalous node without worsening predictions for unseen topologies.  To help the models learn, we augment node features with node indices (one-hot), so that $d = n+1$. 

\subsection{Evaluation results}\label{sec:synth:results}

\subsubsection{Simple-sum toy case}
\begin{figure}[t]
\centering
\includegraphics[trim=11pt 22pt 11pt 10pt,clip, width=0.9\columnwidth]{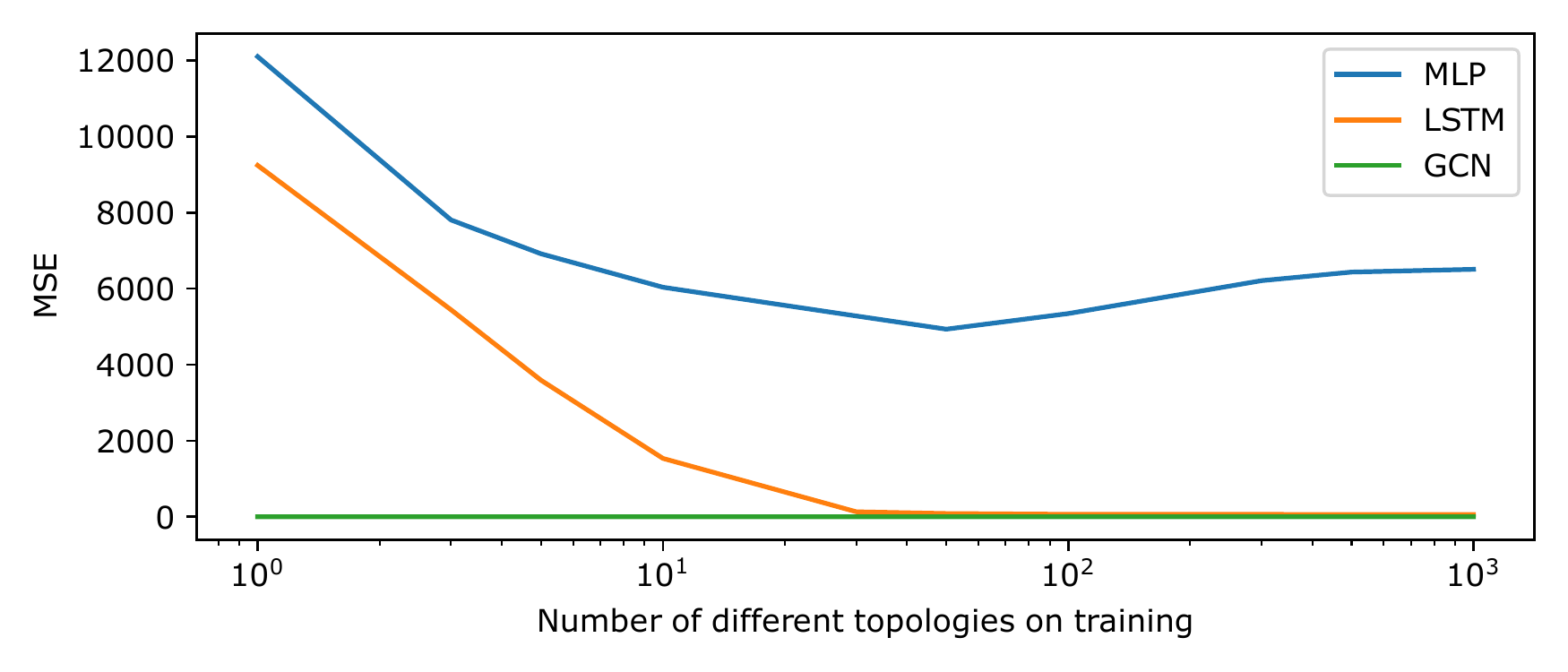}
\includegraphics[trim=11pt 10pt 11pt 10pt, clip, width=0.9\columnwidth]{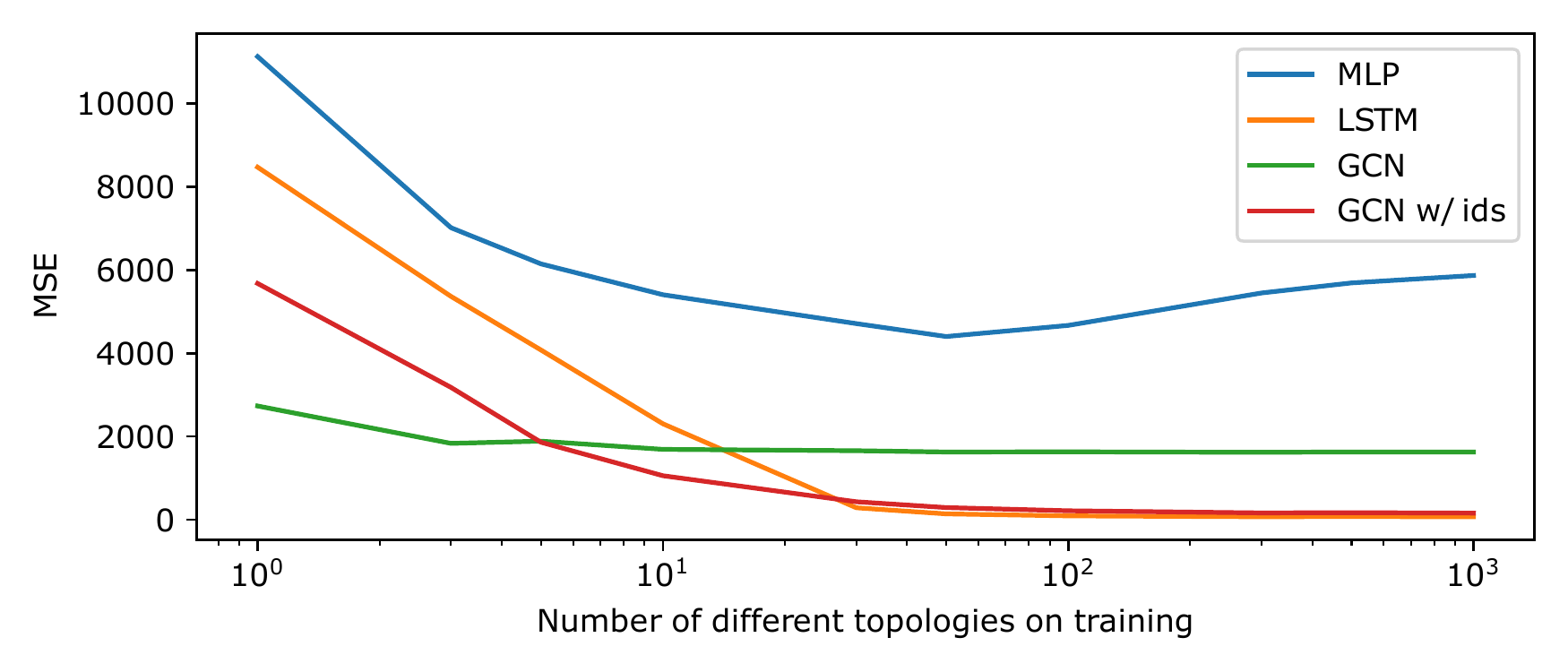}
\caption{MSE error as a function of the number of different possible topologies $k$ on training for simple sum (top) and single AP failure (bottom) scenarios.}\label{fig:synth}
\end{figure}

We first start with the simple sum toy case, i.e.\ the true interference is the sum of neighbor loads. Fig.~\ref{fig:synth} (top) shows the obtained results for 10-node graphs. As expected, the 
GCN always attains 0 loss. The MLP has the worst predictions, and the LSTM needs to see many different topologies in order to attain a small loss.   This suggests, as expected, that the GCN is the preferred option for building  models that transfer across topologies, without needing to be trained on many networks.

\subsubsection{Single AP failure}
Bottom plot of Fig.~\ref{fig:synth}  shows the results for the second problem where the picture gets  more complex. 
We observe that LSTM can natively learn the misbehavior of certain particular nodes, given enough variety in network topologies, while  GCNs cannot. A counter measure is to augment the GCN with node indices.

\begin{figure}[t]
\centering
\includegraphics[trim=30pt 20pt 40pt 10pt, width=0.7\columnwidth]{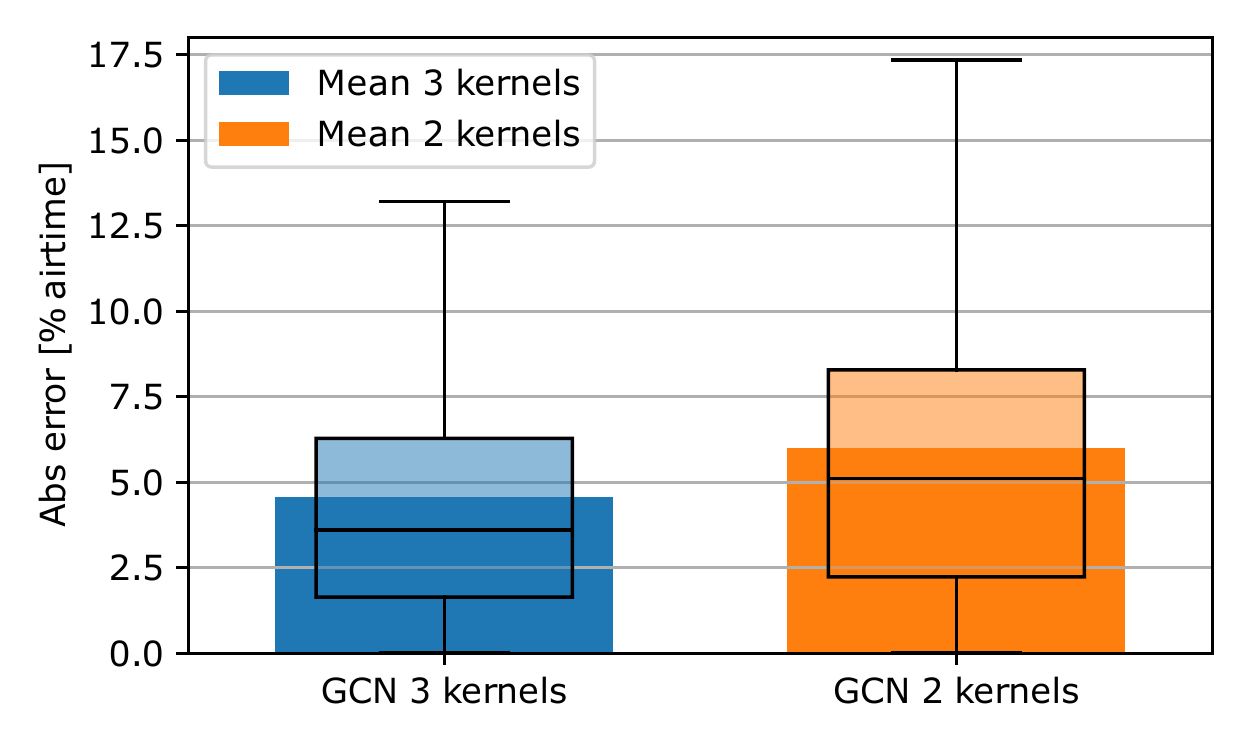}
\caption{Ablation study for kernels}\label{fig:problem_kernels}
\end{figure}

\subsubsection{Impact of kernels}\label{sec:kernels}
Moving away from our synthetic scenarios, we now verify that real RSSI measurement actually help in the interference estimation. On $70\%$ of samples from the \Nseven\ network, we train one GCN with kernels $T_1$ and $T_2$ (\emph{2 kernels}) and one also including $T_3$ (see Sec.~\ref{sec:GCN}), showing in Fig.~\ref{fig:problem_kernels} the results on the remaining 30\%, displaying  the mean absolute error (MAE) across samples, the median, quartiles 1 and 3 (boxes), and 5th and 95th percentiles (whiskers). Observe that the use of the third kernel reduces the MAE by $33\%$ (from $6\%$  to $4\%$), thus we use 3 kernels from here on.

\section{Real network application}
\label{sec:real}
\begin{figure}[t]
\centering
\includegraphics[trim=30pt 20pt 40pt 0pt, width=0.8\columnwidth]{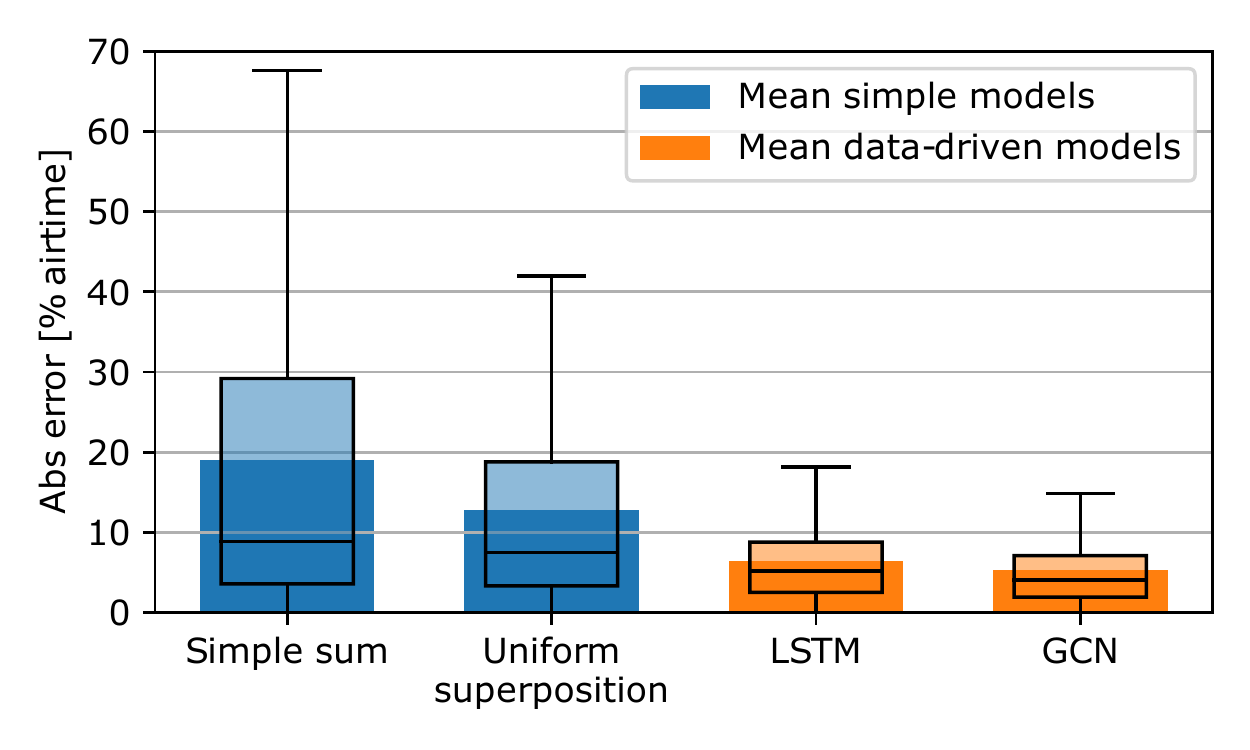}
\caption{Performance on \Nseven\ network.}\label{fig:samenetwork}
\end{figure}

After this initial study of neural architectures for our problem, we now move to apply them to real networks as an illustration. This section trains the interference models on real data from the \Nseven\ network, while testing  performance  both on \Nseven\  and \Nsix. We only regard samples with high-load (interference $\geq 10\%$ of airtime) out of three reasons: (1) Low-load/low-interference scenarios are not of interest in practice since network performance is barely affected in this interference region. (2) Noise is more prevalent in the low-load regime. (3) It allows a fair comparison between \Nseven\ and \Nsix. Thus, we oversample high-load samples by factor 10 during training, and only test on high-load samples.
Each time,  we compare the models against the simple sum and superposition models as a baseline. 

We  start with the case where the models are trained and valiadated on the same network \Nseven. 
Fig.~\ref{fig:samenetwork} shows that both models  outperform the simple estimations. In particular, the MAE decreases from 18\% (simple sum) and 13\% (uniform superposition) to 5\% (LSTM) and 4\% (GCN). The quartiles, shown by the boxes, and the 5th and 95th percentiles (whiskers) confirm the order among the models.

\begin{figure}[t]
\centering
\includegraphics[trim={0pt 0pt 0pt 0pt},clip, width=1\columnwidth]{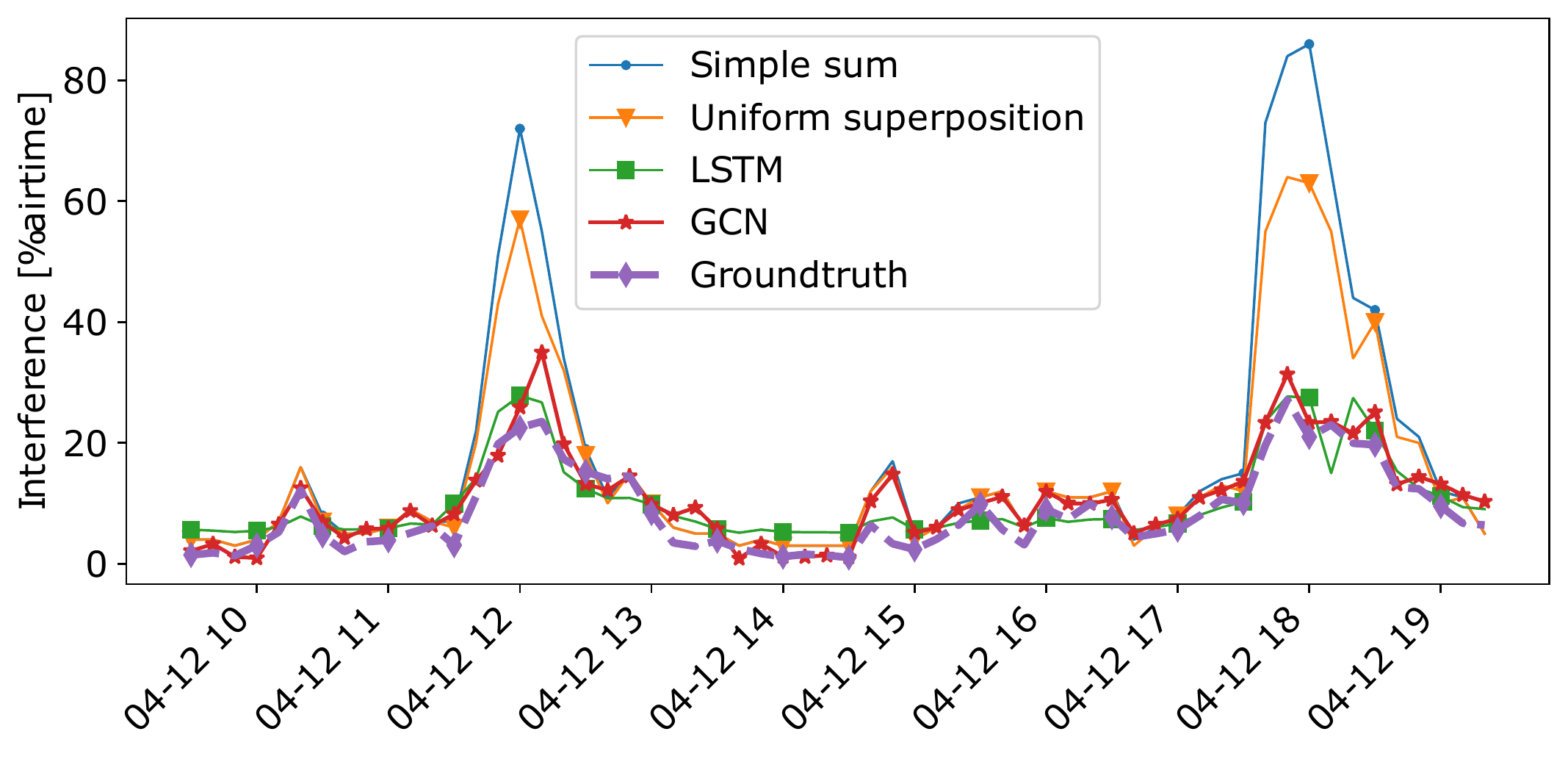}
\caption{Predictions for one AP, one day.}\label{fig:illustration}
\end{figure}

For intuition, we show in Fig.~\ref{fig:illustration} an example of the estimated interference over time for one AP on one day, compared to the groundtruth and to the simple estimation methods. Observe that here both simple models overestimate interference particularly during high-load periods, while both LSTM and GCN provide a much more precise estimation.

\begin{figure}[t]
\centering
\includegraphics[trim=10pt 12pt 10pt 11pt, clip,width=1\columnwidth]{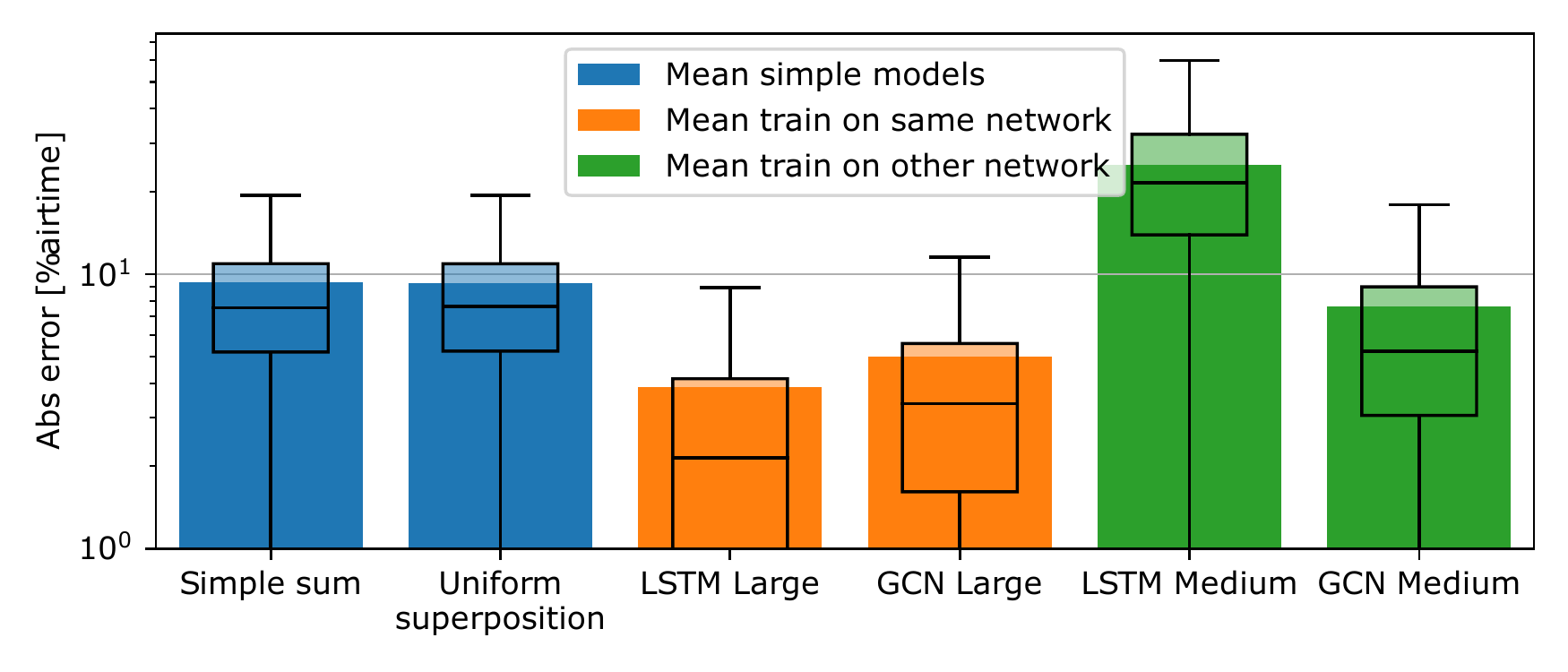}
\caption{Results for \Nsix\ network including transfer.}\label{fig:differentnetwork:all}
\end{figure}

Now, we evaluate the performance of models trained on the \Nseven\ network when applied to the \Nsix\ one.\footnote{Note that the input to the neural network needs to be padded to enable the transfer to other networks.} Fig.~\ref{fig:differentnetwork:all} shows firstly that the simple models provide a better estimate here than on \Nseven\ (both at MAE 9\%), and both deliver very similar results. The latter fact indicates that concurrent transmissions are not as prevalent on \Nsix\ as on \Nseven. Secondly, LSTM trained on \Nsix\ (LSTM Large) outperforms GCN Large (5\% airtime vs 4\%). This may be due to the fact that the GCN does not see enough diversity of high-load samples during training. 
Thirdly, the GCN trained on \Nseven\ (GCN Medium) provides better interference estimations than the simple models (7.5\% MAE vs 9\%) while LSTM Medium fails to transfer (25\%). We expect that training the GCN on few more networks will lower the error. We plan to extend the training as part of our future work.


\section{Related Work}
\label{sec:related}
Prior work estimated different network metrics and events. The most prominent ones treat throughput prediction \cite{ soto2021atari, lin2009machine, kajita2014channel},  RSSI for handover prediction \cite{khan2022ml, park2013handoff, montavont2006ieee}, interference \cite{shrivastava2011pie}, and AP selection based on other factors \cite{khan2022ml, vasudevan2005facilitating, bejerano2004fairness}. 
In last two  cases, the studies rely on past measurements to predict future values, designing the problem as an auto-regressive process \cite{park2013handoff}. 
The authors of \cite{shrivastava2011pie} predict \emph{physical} WLAN interference based on fine-grained packet-level data. Similarly, authors of \cite{jimenez2017non} aim at predicting WLAN QoE metrics from  WLAN frames. Such  approach is not suitable for what-if scenarios that we are aiming for. 
A particularly interesting recent work~\cite{soto2021atari}, closer to ours  approach, predicts user throughput leveraging GCNs, using node features like adjacency, channel configuration and device type to predict throughput.   


\section{Conclusions}\label{sec:conclusions}
In this paper, we methodically tackled the WLAN interference estimation problem, gathering the following observations:
\noindent (i)  WLAN networks confirm their reputation of being ``unpredictable'': Neighbor relations change over time and certain outlier nodes consistently exhibit anomalous behavior.
\noindent  (ii) Machine learning models outperform simple sum estimation but come at different costs. As shown in synthetic settings, LSTMs need to be trained on many networks to generalize, but they can natively learn the misbehavior of particular anomalous nodes.  Unless explicitly provided with node indices during training, GCNs do not allow this, off-the-shelf, but they can better transfer compared to their  LSTM counterpart.
\noindent (iii) Even when trained on a single real network of 33 AP nodes, our GCN model still outperformed simple models in predicting interference in another larger network of 84 APs. While this of course needs to be confirmed on other networks, the ability to apply the model as-is without retraining lowers the barrier of adoption.

\bibliographystyle{ACM-Reference-Format}
\bibliography{ref_short}


\end{document}